\documentclass[12pt]{article}
\usepackage[english]{babel}
\usepackage{amssymb,graphicx,wrapfig}
\author{Vladimir A. Petrov\footnote{Vladimir.Petrov@ihep.ru}
 and Nikolai P. Tkachenko\footnote{Nikolai.Tkachenko@ihep.ru}}
\index{}
\title{On supposed oscillations of differential 
cross sections in $ pp $ - scattering at $ \sqrt{s} = $13 TeV.}
\date{}

\begin{document}

\maketitle
\begin{center}

A.A. Logunov Institute for High Energy Physics 

NRC "Kurchatov Institute", Protvino, RF
\end{center}
\begin{abstract}
The question of possible existence of oscillations in the region of diffraction peak in $pp$ - scattering at
$\sqrt{s}=13$ TeV is considered in detail. It is shown that, within
available experimental data published by the TOTEM and
ALFA/ATLAS, posing the question of searching for such a subtle effect looks premature.\vspace{-3.6mm}
\end{abstract}

\section*{Introduction\vspace{-3.6mm}}
It is well known that diffraction scattering of hadrons exhibits properties similar to
optical diffraction pattern of light scattering off an obstacle~-- for example,
alternating light (maximum) and dark (minimum) areas on the screen. In hadronic
scattering a similar structure also reveals , namely, {\it dip}, the first minimum,
next to the diffraction peak. In principle, with larger transferred momenta $ \vert t \vert $, 
other diffraction maxima/\\ minimums are possible. The location of these
structures is associated with the interaction region in the space of impact parameters and
is determined by the interference of alternating contributions from the eikonal (the “Born amplitude”).

At various times, results were published and discussed in which it was stated that, in addition to
the above diffraction structures, in particular in the peak region, some less pronounced, but statistically
significant structures, {\it oscillations}, modulating the $ t $-dependence of differential cross sections were
observed. Such phenomena would indicate that our ideas about the mechanism of hadron scattering
need a serious revision \cite{Sel}. 

Let us notice that the known theoretical models do not predict any fine structure of the angular
dependence of scattered particles\footnote{Some specific arguments for the absence of oscillations at high
enough energies can be found in Ref.\cite{Tr}}. 

Note that in ref. \cite{Sel} and \cite{Graf} the work  \cite{Mart} is mentioned (as an example of the
oscillations in question), in which, based on the general principles of quantum field theory, it was shown
that for some types of amplitudes the so-called “AKM scaling” takes place , which consists in the existence
of a limit
\[\lim _{s\rightarrow\infty, \tau fix} \frac{T_{N}(s,t)}{T_{N}(s,0)}= f(\tau).\]
Moreover, the function $ f(\tau) $ is an entire
a function of order 1/2, oscillating in the neighborhood of the real axis, for example, like
$J_{1}(\sqrt{\tau})/\sqrt{\tau}$. However, these “oscillations” are nothing more than asymptotic images of
ordinary diffraction structures, which do not indicate any new physical effects. “AKM scaling” itself refers to
the “ultra asymptotic” energy region and is rather some asymptotic boundary condition for checking the
compliance of one or another models to the general principles of quantum field theory. In this sense, the
properties of the function $ f(\tau)$ do not have a direct relationship to the region of real energies
(including, to a large extent, the LHC).

Of course, the conclusion about the presence (or absence!) of oscillations should be based on a detailed
statistical analysis, as a result of which the oscillations are clearly identified (or ruled out) within the
framework of the available experimental data.

In this work, we undertook such an analysis using the example of experimental data on $pp$-scattering at
$\sqrt{s}=13$ TeV\footnote {Regarding oscillations at energies of 2.76, 7 and 8 TeV  It was noted in
\cite{Graf} that for the discussion of the issue of oscillations at 13 TeV, these results are not statistically
significant.}.
At this energy the following differential cross section measurements are available: two by
collaboration TOTEM and one by group ALFA/ATLAS
\footnote{In what follows we will use
only the term "ATLAS"}.
 
A favourable circumstance was the fact that these three data sets have a common interval:
$0.035 \lesssim |t| \lesssim 0.2$ GeV$^{2}$, which makes a comparison of the results on this interval
justified  for all three arrays of experimental data.

As a result of the analysis, we came to the conclusion that with the existing quality the experimental data
cannot lead to any definite conclusion regarding the presence of oscillations in the differential cross section
of $pp$-scattering at $\sqrt{s}=13$ TeV.

To analyze oscillations  it is necessary to choose a phenomenological model (as a “background”)
which describes the experimental data quite well. There are many such models, especially in the field of
small momentum transfers. As a basis, we used the model \cite{Martynov}, which, after a proper
modification, made it possible to achieve a record value of $\chi^2/$DoF for the entire set of experimental
data at $\sqrt{s}>7$ GeV and $|t |<5$ GeV$^{2}$. These results are published in the 2014 and 2016
editions of PDG for an early version of this model.

Of course, as shown in \cite{Ptr}, conceptually,  this model cannot
claim physical significance. In addition, it also contains quite a large
number of parameters. However, as a descriptive tool that well parameterizes the background, it best
matches the experimental data to the widest possible extent.
areas of energy and momentum transfer. Let us note here that this model also did not allow us to obtain a
statistically reliable description of hadronic diffraction data: the record value of $\chi^2$/DoF $\cong$ 1.8
corresponds to an extremely low level of confidence. However, this the situation may well be improved, as
we will present arguments at the end of the article.\vspace{-4.6mm}

\section{Experimental data from ATLAS and TOTEM\vspace{-2.6mm}}
ATLAS experimental data \cite{ATLAS} are presented in full not only in
the printed work, but also in the HEP DATA database, including covariance matrices of statistical and
systematic errors, which, unfortunately, cannot be said about the measurement results from the
TOTEM collaboration.
\begin{figure}[htb]
$$
\includegraphics[width=165mm]{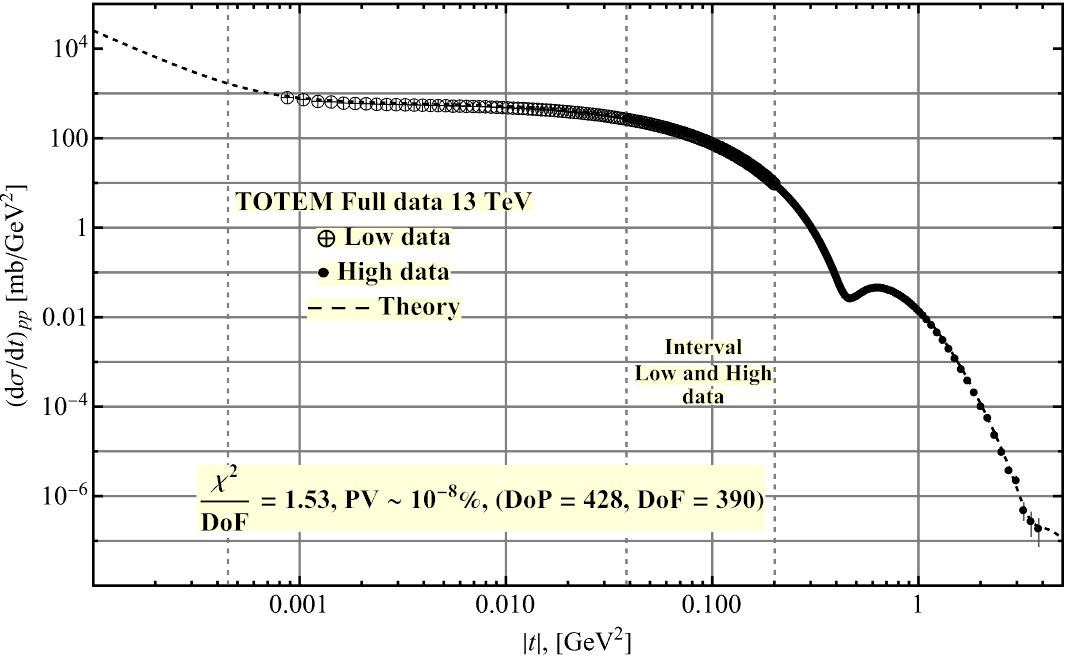}\vspace{-3.1mm}
$$
\parbox{190mm}
{\caption{
$(d\sigma /dt)_{pp}$ {\it from two} TOTEM {\it measurements. Crosses denote the Low data array while
the dots represent the High data array. The total errors of these measurements are virtually
indistinguishable in this figure, except for the three points with the largest} $|t|$ {\it values. The data from
these two measurements intersect in the region} $|t|\in\backsim [0.035 \div 0.2]~\mbox{GeV}^2$.
{\it The dotted curve is the result of a joint fit of Low and High data and is shown for illustration. Visually it
describes everything perfect but has zero confidence level.}
}
}\vspace{-4.1mm}
\label{pic1}
\end{figure}
TOTEM data published its two differential cross-section measurements in
\cite{TOTEM1,TOTEM2}. In this case, in one version the sources of systematic errors are given, which
makes it possible to construct a systematics covariance matrix. However, for statistical errors
there is no such data. In this case, only complete systematic and
statistical errors are in use, and there is no talk of covariance matrices at all.

Note that two arrays of TOTEM measurements intersect on the interval
$|t|\in\backsim [0.035 \div 0.2]~\mbox{GeV}^2$~-- see pic. \ref{pic1}. At the intersection of the
experimental points, we actually have two independent measurements of the differential cross section at
the same accelerator and at the same energy, which is a favourable circumstance for further comparison
of the results.

Below the experimental TOTEM points measured at lower values are presented.
We will denote points with low $|t|$ as \textbf{Low data}, while with the higher values  \textbf{High data}.
The union of these two arrays is denoted as  \textbf{Full data}.

Fig. \ref{pic1} shows experimental results $d{\sigma}/dt_{pp}$,
measured at the TOTEM installation at energy $\sqrt{s}= 13$ TeV\footnote{Further we will not constantly
mention the energy of 13 TeV, because only this energy is considered in this work.}. Experimental ATLAS
data are not shown in this graph, because on this (logarithmic) scale they are poorly distinguishable from
the TOTEM data. Next we present graphical results of the ATLAS experiment when it is visually
comfortable.

According to the recommendations of the ATLAS authors, we do not use two experimental points with
minimum values of $ \mid t \mid $. Then it is implied everywhere, although we show these two points on
the graphs. Thus, further in all fits we use only experimental ATLAS points for
$|t|>0.00045~\mbox{GeV}^2$.

Experimental ATLAS data are preferred for analysis. They allow one to compose $\chi^2$ functions taking
into account correlations between points, with use both of systematic and statistical errors. Unfortunately,
experimental TOTEM data do not provide such a possibility. Therefore, we can compare these results in
the same (homogeneous) way only by composing the function $\chi^2$ without taking these correlations
into account and with help of the standard formula: 
\begin{equation}
\chi^2 = \sum_{i}
\left(
\frac{theory(experiment_{i})-experiment_{i}}{error_{i}}
\right)^2\vspace{-2.6mm}
\label{eq1}
\end{equation}
\newpage
~\vspace{-11.1mm}

\section{TOTEM\vspace{-2.6mm}}

\subsection{Fit with complete errors\vspace{-2.1mm}}
Fitting results when composing the function $\chi^2$ with use of complete errors
$\mbox{Err}_{full}= \sqrt{\mbox{Err}_{stat}^2+\mbox{Err}_{syst} ^2}$ are as follows:\vspace{-2.1mm}
\begin{enumerate}
\item For \textbf{Full data} $\chi^2=137.033$, DoF = 390, which corresponds to a low confidence level.
        For this reason, we will not consider this fit result further.\vspace{-2.1mm}
\item Fits with complete errors for each of the two TOTEM data sets also give a zero level of
        confidence.\vspace{-2.1mm}
\end{enumerate}
Thus, we can only use fitting with account of only statistical errors.\vspace{-2.1mm}

\subsection{Fit with statistical errors\vspace{-2.6mm}}
\begin{enumerate}
\item {\bf Full data}: $\chi^2_{\Sigma}=595.36$, DoF = 390, which corresponds to a zero confidence
        level. For this reason, we do not take this option into consideration.\vspace{-2.1mm}
\item {\bf Low data}: $\chi^2 =123.2,~\mbox{DoF} = 100,~ \mbox{CL} = 5.77$\%\vspace{-2.1mm}
\item {\bf High data}: $\chi^2 =232.8,~\mbox{DoF} = 252,~ \mbox{CL} = 80.2$\%\vspace{-2.1mm}
\end{enumerate}

Actually, only the last two results can be relied upon to some extent, due to their significant level of
confidence (especially for the \textbf{High data} array).
We have no other results with significant confidence levels when constructing $\chi^2$ using the formula
(\ref{eq1}).

\begin{figure}[h]
\noindent
$$
\includegraphics[width=155mm]{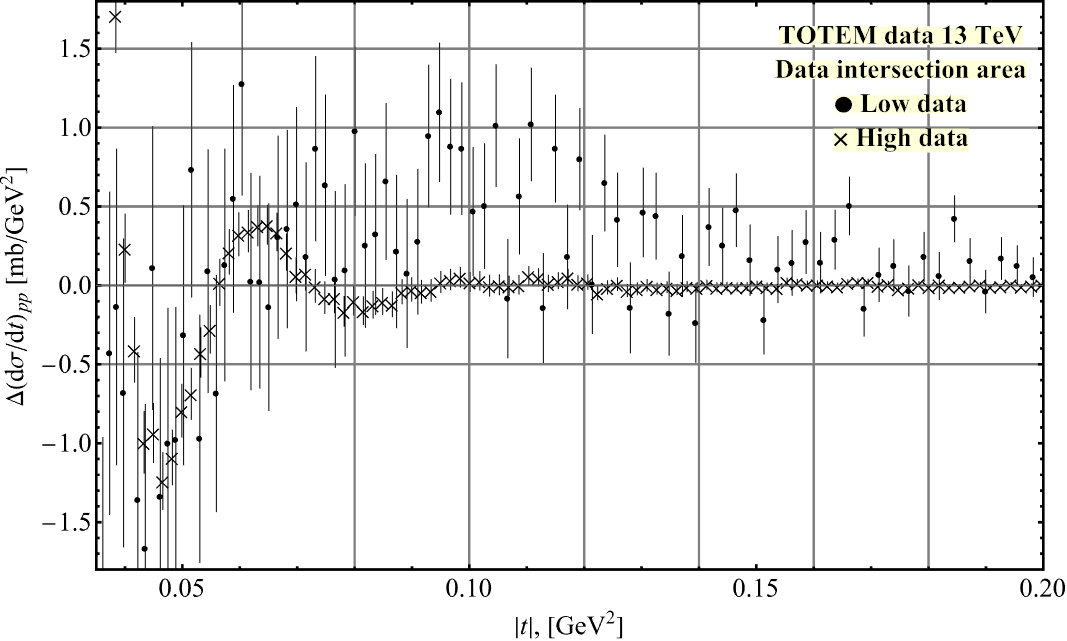}\vspace{-2.1mm}
$$
\parbox{190mm}{\caption{$(d\sigma /dt)_{pp}^{\mbox{experiment}}-
(d\sigma /dt)_{pp}^{\mbox{theory}}$ {\it for the \textbf{Low} and \textbf{High data}} TOTEM.
{\it Crosses denote the \textbf{High data} array, and dots the\textbf{Low data}. The interval where both
these arrays intersect is given.}}\vspace{-3.1mm}
\label{pic2}}
\end{figure}

Let us now consider the deviations of the experimental points from the theoretical curve for the 2nd and
3rd cases:\vspace{-2.1mm}
$$
\Delta (d\sigma /dt)= d\sigma /dt_{experiment}(t_i)-d\sigma /dt_{theory}(t_i)\vspace{-1.1mm}
$$
These results are presented in Fig.\ref{pic2}.

\newpage
~\vspace{-7.1mm}

Experimental \textbf{Low data }points (points on the graph) do not show a tendency to group into an
oscillatory process, while the\textbf{High data} points (crosses on the graphs) clearly allow description by a
damped sinusoid, at least in the interval $0.04 \lessapprox |t|\lessapprox 0.12\ mbox{GeV}^2$. For a
model description of these points, one can write out as many damped sinusoids as desired with an
excellent level of confidence, which is a consequence of large (statistical) errors for these difference points.
Moreover, even just a zero straight line gives an excellent level trust. 

Thus, despite the fact that the central values  of the array clearly gravitate toward oscillatory behaviour, it
is not statistically possible to conclude that it exists.

Note that the Low data points, {\bf which can be considered as a repeated measurement for the High
data points}, show no signs of oscillatory behaviour at all. But even if such behaviour did occur, it would be
impossible to conclude that there are fluctuations for reasons similar to those indicated when describing
the \textbf{High data }array.

For completeness, we note that the oscillatory process for the TOTEM data looks much clearer than in
Fig.\ref{pic2}, if we consider possible oscillations on the full TOTEM data array (\textbf{Full data}) and on
the same area of $t$. However, since these possible fluctuations appear in relation to the fit with zero
confidence, then we also discard this option.

\begin{figure}[htb]
\noindent
$$
\includegraphics[width=155mm]{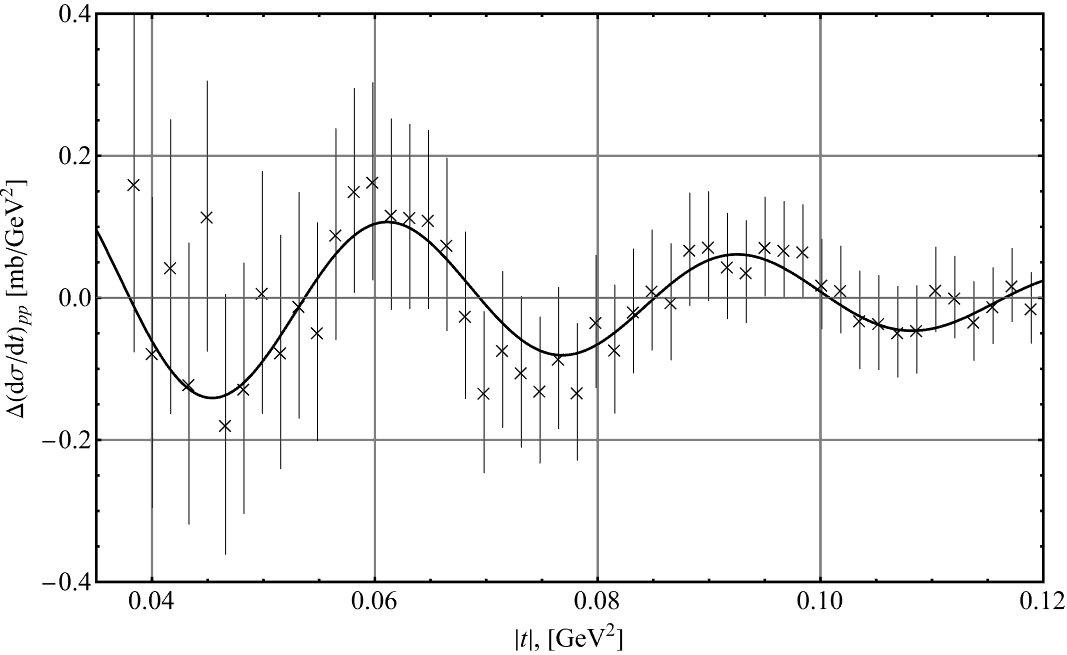}\vspace{-3.1mm}
$$
\parbox{180mm}{\caption{{\it Possible oscillatory process on the \textbf{High data} array in on an 
enlarged scale (see Fig. \ref{pic2}). A solid curve is one of the possible ways to describe this process, but
since an infinite number of such curves can be drawn, this curve has purely illustrative value. A possible
oscillatory process manifests itself at a level not exceeding the experimental statistical error.}}
\label{pic21}}\vspace{-2.1mm}
\end{figure}

Thus, we have the only option with suspicion of possible oscillations. Therefore, we present this graph on
an enlarged scale - Fig. \ref{pic21}.

The theoretical curve for the \textbf{Full data} array, as we saw above, visually passes well across all
experimental points, but is not statistically reliable. The description of the \textbf{Low data} array cannot
be considered statistically reliable either. (confidence less than 6\%). So only the description of the
\textbf{High data} array is statistically reliable. However, to connect these two arrays of measurements
obtained at the same facility and at the same energy, to combine them in a single description purely is
impossible for statistical reasons. These two arrays of measurements and their combination lead to other
contradictions when trying to use them as a model.

\begin{figure}[htb]
\noindent
$$
\includegraphics[width=175mm]{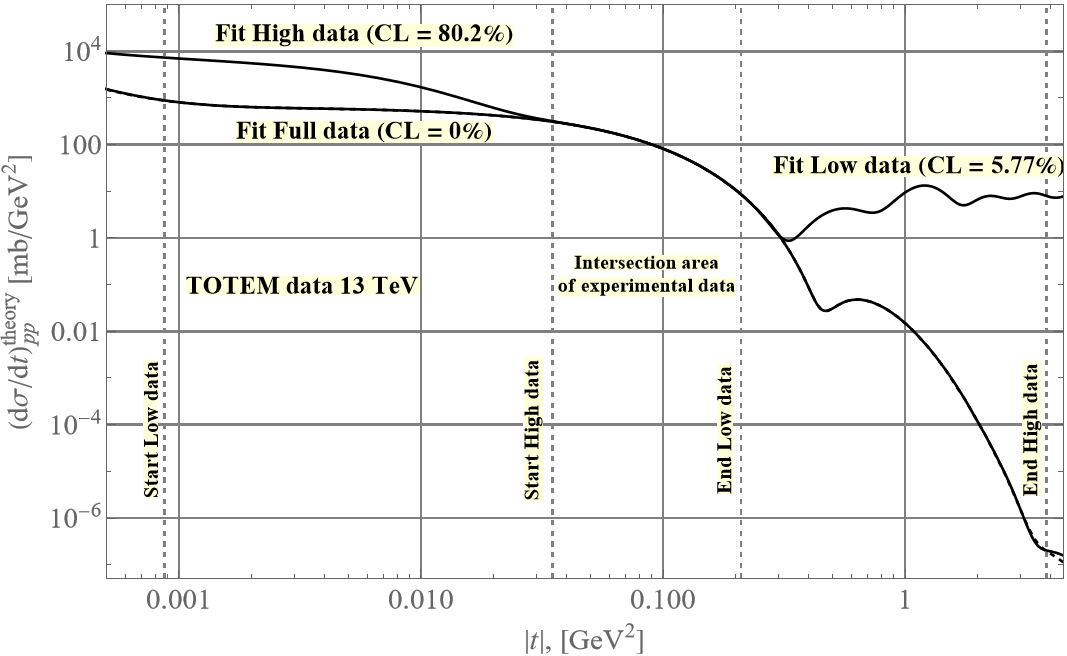}\vspace{-3.1mm}
$$
\parbox{190mm}{\caption{{\it Model curves corresponding to the fits of \textbf{Full data, Low data }and 
\textbf{High data}. Experimental points are not given, since on this scale they are visually
indistinguishable from the given curves.}}\vspace{-3.1mm}
\label{pic3}}
\end{figure}

The model used here describes the \textbf{Low data} array more or less satisfactorily (confidence ~6\%),
but outside of this array it behaves in a completely unacceptable way, Fig\ref{pic3}. In the common
interval with the \textbf{High data} array, it visually completely coincides with the curves
obtained on the \textbf{Full }and \textbf{High data}. In addition, the fitting curve for the \textbf{Low
data} describes well the region of experimental data with small $|t|$ (visually it coincides with the
\textbf{Full data}).

On the contrary, the curve obtained from the\textbf{High data} describes well experimental points with
large $|t|$ (coincides with the curve describing the \textbf{Full data}) and is completely unsuitable for
describing experimental data with small values of $|t|$.

Of course, all three of these curves do not have to exactly repeat each other. But even their existing
qualitative behaviour over the entire range of experimental data is unsatisfactory and completely
unacceptable. These contradictions manifest themselves in other ways as well. For example, the values of
some parameters obtained from fits of two experimental arrays experience an unacceptably huge,
sometimes order of magnitude, discrepancy between each other. So, for example, the parameter that is a
factor of the squared logarithm in the expression for the total cross section is in one case equal to
0.224 mb\footnote{This is in a good agreement with the processing of our experimental data on total cross
sections, repeatedly published in PDG.}, whereas in another case this same coefficient increases by more
than an order of magnitude. In addition, in both cases, some parameters acquire enormous, physically
absurd values. 
\newpage
~\vspace{-9.1mm}

\section{ATLAS\vspace{-2.1mm}}

\begin{enumerate}
\item $\chi^2 = 1.157$, DoF = 40, CL = 0 when fitting with complete errors. This corresponds to a zero
        confidence level and therefore we exclude this option from consideration.\vspace{-2.1mm}
\item $\chi^2 = 69.63$, DoF = 40, CL = 0.25\%, when fitting only with statistical errors. This option also
        has a very low level of reliability and for this reason we also exclude it
        from consideration\vspace{-2.1mm}
\end{enumerate}

Thus, we cannot identify the supposed oscillations in experimental data due to the lack of a reliable model
description of them. The most likely reason for this situation, in our opinion, is, as in the case of TOTEM, a
shift in the experimental data of the differential cross section from their true value. Below we provide a
preliminary study of this issue.\vspace{-6.1mm}

\section{On a  possible shift of experimental data on differential sections\vspace{-3.1mm}}
\begin{figure}[htb]
\parbox{130mm}{\includegraphics[width=125mm]{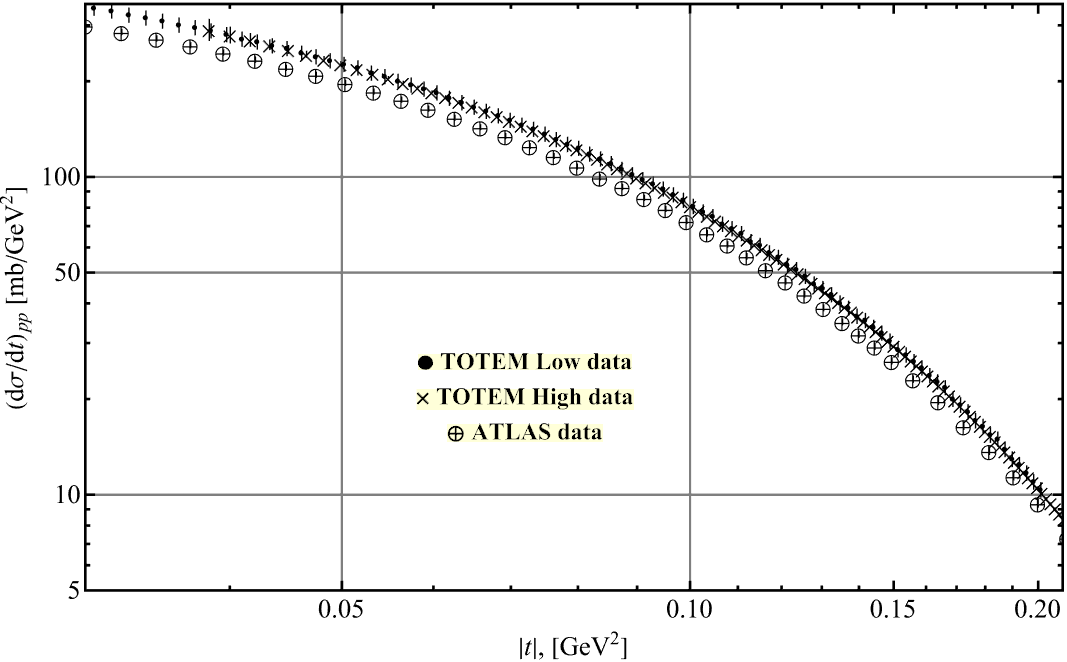}}\vspace{0.1mm}

\parbox{130mm}{\includegraphics[width=125mm]{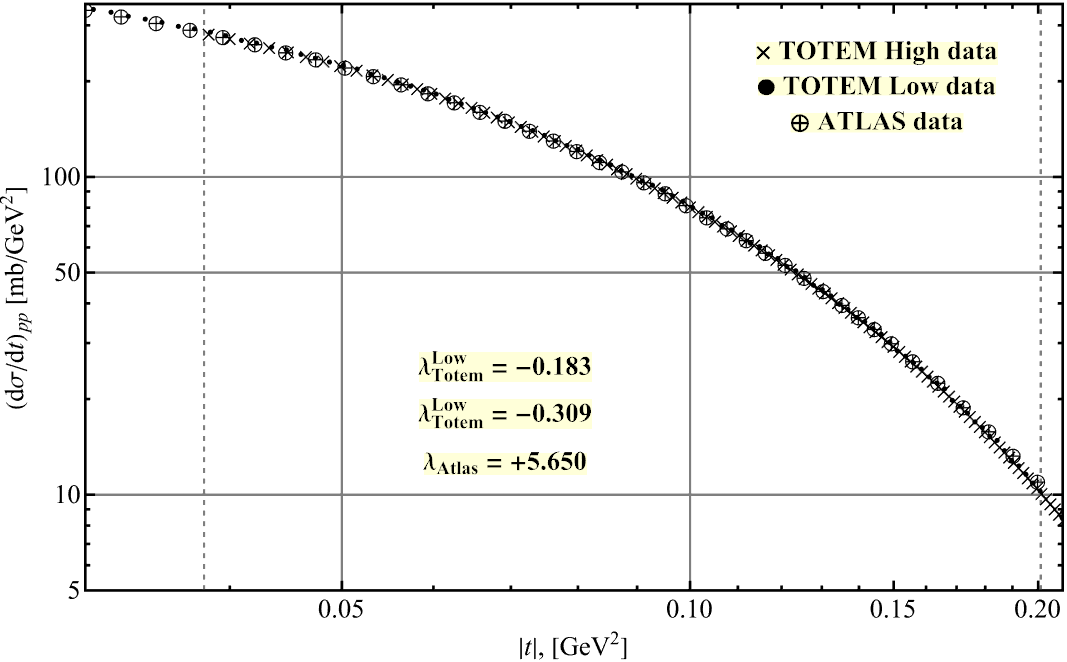}}\vspace{-2.1mm}

\parbox{180mm}{\caption{{\it 
On the top are the original experimental data} TOTEM{\it and} ATLAS {\it depending on the transferred
momentum in the interval where both of these arrays intersect. Statistical errors are indistinguishable at
this scale. Below is the same thing, but with the central values of the data shifted. Data} TOTEM {\it shifts
slightly downward - each point by value} $\lambda^{Low}_{TOTEM}=-0.183$ {\it and}
$\lambda^{High}_{TOTEM}=-0.309$ {\it its systematic error, and the ATLAS points shift significantly
upward with a similar coefficient equal to} $\lambda_{ATLAS}=5.650$.}\label{pic5}}\vspace{-2.1mm}
\end{figure}

Actually, the fact that the experimental points of differential cross sections can be shifted relative to their
true values follows from the general form of the experimental data of TOTEM and ATLAS at $\sqrt{s}=13$
TeV. The center points of these two sets are several standards apart, calling into question the compatibility
of these experimental data. This difference is clearly visible in the top graph of Fig.\ref{pic5}.

The joint fit of complete sets of experimental data leads to an unphysically large value of
$\chi^2$/DoF > 100, which corresponds to a low confidence in the result, and this is a quantitative
argument in favour of the fact of the incompatibility of these data.
We made an attempt to artificially combine these data, artificially shifting their experimental central values.
One of the possible ways to shift experimental points is as follows: we shift each central value of the
experimental TOTEM value by an amount proportional to the systematic error of this point.

Let us denote the proportionality coefficients for each measurement at TOTEM $\lambda^{Low}_{TOTEM}
$ and $\lambda^{High}_{TOTEM}$. It is not a priori obvious that they will be the same, although visually
it appears that way. Let us subject the central values of the ATLAS  experiment data to a similar shift with a
coefficient $\lambda_{ATLAS}$ that is the same for all ATLAS experimental points. Further, let us compose
the function $\chi^2$ with new central values, while setting the coefficients $\lambda^{Low}_{TOTEM}$,
$\lambda^{High}_{TOTEM}$ and $\lambda_{ATLAS}$ with additional fitting parameters\footnote{In the
expression for $chi^2$, three additional penalty functions will appear due to the presence of new
degrees of freedom $\left(\lambda^{High}_{TOTEM}\right)^2$,
$\left(\lambda^{Low}_{TOTEM}\right)^2$ $\left(\lambda_{ATLAS}\right)^2$,
appeared as a result of the introduction of three new parameters.}. The results of such a fit are shown in
the bottom graph of Fig. \ref{pic5}.

It is striking that the two TOTEM measurement arrays are also not entirely consistent with each other; to
achieve the best value of $\chi^2$ they are shifted by different amounts.

However, the fundamental thing in this option is the fact that even artificial shifts of experimental points do
not lead to a statistically significant result: $\chi^2 = 2312.82$, DoF =
$N_{\mbox{tot}}-N_{\mbox{param }}=506-41=$365. CL = 0. Thus, it is also not possible to talk about
possible oscillations even at artificially shifted experimental points.

However, it is important that the shift in experimental data sharply reduces the value of $\chi^2$. At the
same time, it is not at all excluded that in other, earlier experiments, such a shift in experimental data
takes place. Therefore, when joint global data processing is necessary, after the initial data processing, it is
necessary to fix the obtained parameters and fit all experiments using the data shift procedure described
above, which will certainly lead to a decrease in $\chi^2$.

Next, one should record this new (shifted) data and re-run the fit to determine the model parameters
again. Then, one should repeat this procedure as many times as necessary until $\chi^2$ is fixed at some
value. This will lead to a decrease in the total $\chi^2$, possibly to statistically significant values. Now the
experimental data from approximately 6000 points shows a record value of $\chi^2$/DoF $\simeq 3$,
which is statistically completely unacceptable in terms of the level of confidence. Thus, there is still no
general description of the entire set of accumulated data on proton-proton scattering. Of course, changing
the experimental data is not a very welcome procedure, but there is no other choice. Moreover, if no new
experimental data are expected.

However, in our opinion, the model also needs to be improved: the current, albeit best, description of
the experimental results has too many parameters, not to mention other problems
\cite{Ptr}. 
\newpage
~\vspace{-11.1mm}

\section{ATLAS data taking into account correlations}\vspace{-2.1mm}

In conclusion, we will move away from the method of describing differential  cross-sections with account of
statistical errors only. Given the high quality of presentation of experimental data in the ATLAS experiment,
we will conduct the study taking into account correlations between experimental points, using weight
matrices \footnote{Two points with minimum values of $ \mid t \mid $ are still discarded in accordance
with the recommendation of the authors of the ATLAS publication.}.

\begin{figure}[htb]
\parbox{130mm}{\includegraphics[width=130mm]{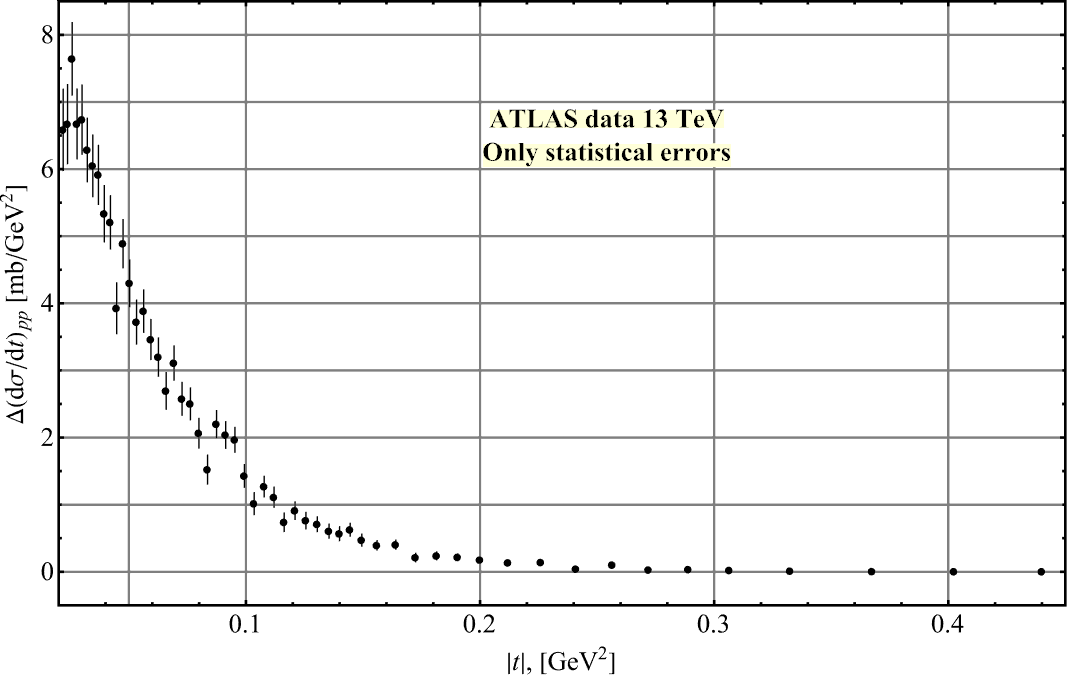}}\vspace{0.1mm}

\parbox{130mm}{\includegraphics[width=130mm]{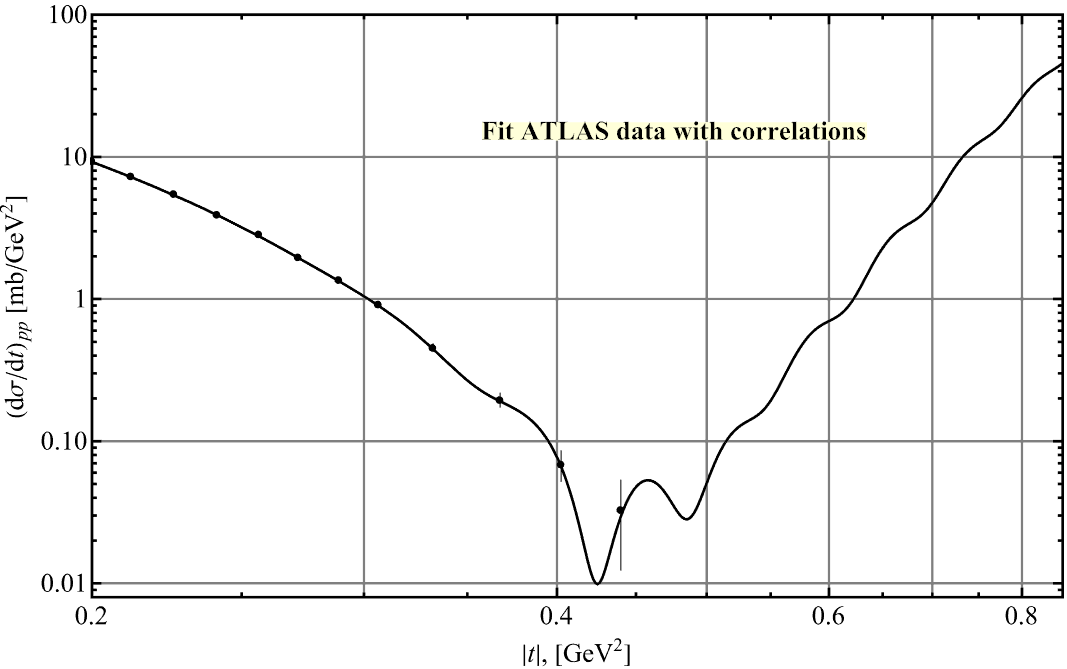}}\vspace{-0.1mm}

\parbox{190mm}{
\caption{Top graph:{\it Deviation of ATLAS experimental points from the model curve
for fit with weight matrix. Total errors are significantly larger than statistical ones, and they
go beyond the graph, so they are shown only statistical errors .\newline}
Bottom graph:{\it Behaviour of the model curve in the fit version with a weight matrix
in the region where the experimental points end} ($|t|>0.44$ GeV$^2$).
}\label{pic6}
}\vspace{-4.1mm}
\end{figure}
\newpage
~\vspace{-9.1mm}

Fitting gives the following result:
$\chi^{2}_{\Sigma} = 49.167$, DoF = 40, CL $\simeq$ 22\%\footnote{CL =
$\frac{1}{2^{DoF/2} \Gamma (DoF/2)}
\int_{\chi^{2}_{\Sigma}}^\infty z^{DoF/2-1} e^{-z/2} dz$}. 
This is a fairly noticeable level of confidence, and it is significantly better than when fitting without taking
into account correlations between points\footnote{It is a great regret that it is impossible to compile a full-
fledged weight matrix  from the TOTEM experimental data.}, which gives a significantly higher degree of
confidence to the ATLAS results. However, if we make up the difference between the model fitting results
and experimental points (which makes it possible to identify possible oscillations), then the errors of these
difference points are several times greater than their central values.

The central points themselves do not show any signs of oscillatory behavior. Therefore, in this case, it is not
possible to come to affirmative conclusions about the presence of oscillations.

For illustration, we present this graph, see fig. \ref{pic6}. It can be seen that almost all experimental
points lie above the model curve, and this difference increases monotonically as $|t|$ decreases. This is
nothing more than a manifestation of the properties of the weight matrix (full), and for this reason it
cannot be considered satisfactory.

The properties of the full weight matrix are manifested - and unsatisfactorily - in the behaviour of the
model curve in the interval where the ATLAS experimental data ends: $|t|>0.44\mbox{ GeV}^2$, which is
demonstrated in the lower graph of Fig. \ref{pic6}.\vspace{-4.1mm}

\section{Conclusion}\vspace{-2.1mm}

Only in one place did we have signs of possible (damped) oscillations with a more or less significant level of
confidence: these are the experimental TOTEM points in the  \textbf{High data}. Alas, the second group
of measurements on the same experimental setup and at the same energy (TOTEM  \textbf{Low data})
does not even approximately confirm this conclusion.

It should be noted that artificially shifted experimental points again demonstrate the possibility of damped
oscillations (at the same more or less acceptable confidence levels) on the same TOTEM \textbf{High New}
data array and on the same interval as on unshifted points, but do not confirm it at all this picture on the
TOTEM data array \textbf{Low New}. Thus, the shifted TOTEM points do not demonstrate repeatability of
the effect.

In principle, other methods of shifting experimental points are possible that differ from the method
discussed above. We checked some of them, but we did not get a satisfactory result in favor of oscillations.
To save space and due to the lack of significant results, we do not present these graphs here.

All the above results are based on the use of a model, which (like any other model) is not the only possible
one and may well be subjected, as noted above, to criticism in conceptual terms. However, in a purely
utilitarian sense, we do not yet have a better model as a means of describing experimental data in the
entire range of energy changes.

In our opinion, four important, but mutually exclusive, conclusions follow from everything above:

output:\vspace{-2.1mm}
\begin{enumerate}
\item The measurements, the results of which are used in our analysis, need significant improvement,
        which is especially clearly seen in the experimental data of TOTEM and ATLAS. This circumstance at
        this stage (and in any case) seems to us defining.\vspace{-2.1mm}
\item The model description may be imperfect and needs fundamental modernization. In particular, four
        dozen fitable parameters does not seem to be a satisfactory option.\vspace{-2.1mm}
\item It is possible that new physical effects may appear that are not taken into account by this
        model\footnote{In this regard, we note that in the work \cite{Graf}, to the standard model of the
        nuclear amplitude, a function taken arbitrarily is added that mimics possible oscillations, i.e. just the
        new effects we are looking for. However, the conclusions are similar to those obtained by us.}.
\item As in our previous work \cite{Tka} we noted for the TOTEM experiment that we especially need to
        pay attention to the measurements of points with small $|t|$. In this work, this statement also
        becomes true for the ATLAS experiment. The results of any fits show that the greatest deviations
        from the theory are exactly these points.\vspace{-4.1mm}
\end{enumerate}

\section*{Acknowledgements\vspace{-2.1mm}}
We express our gratitude to A.K. Likhoded for stimulating discussions.\vspace{-4.1mm}

\section*{Appendix: Formulaic description of the model\vspace{-0.1mm}}
The total amplitude $F(s,t)$, with the help of which all observable quantities are calculated, is defined as
the sum of the nuclear $F^N$ and Coulomb $F^C$ amplitudes:\vspace{-0.1mm}
$$
F(s,t)=F^{N}(s,t)+F^{C}(s,t), ~~~
F_{pp}(s,t)=F_{+}(s,t)+F_{-}(s,t), ~~~
F_{{\bar p}p}(s,t)=F_{+}(s,t)-F_{-}(s,t).
$$
$$
\sigma_{tot}(s)=\frac{\mbox{Im}F^{N}(s,0)}{\sqrt{s(s-4m_{p}^{2})}}, ~~~
\rho(s)=\frac{\mbox{Re}F^{N}(s,0)}{\mbox{Im}F^{N}(s,0)}. ~~~
\frac{d\sigma_{tot}}{dt}(s,t)=
\frac{|F(s,t)|^2}{64\pi (\hbar c)^2 s(s-4m_{p}^{2})}.
$$
Let us pay attention to the fact that the amplitudes have dimensions [$\mbox{mb}\cdot\mbox{GeV}^2$].
$m_{p}= 0.93827$ GeV is the proton mass, $(\hbar c)^2=0.389379$ [mb$\cdot$GeV$^2$] (in $c=1$)
The following notations are used:
$$
z_{t}(s,t)\equiv z_{t}=\frac{t+2s-4m_{p}^2}{4m_{p}^2-t}, \equiv
\frac{2s}{4m_{p}^2-t}-1,~~ z(s,t)=2m_{p}^2 z_{t}(s,t),~~
\zeta (s,t)=\ln (-iz_{t})=\ln (z_{t})-i\frac{\pi}{2}.
$$
\subsubsection*{Nuclear amplitude $F^N$}
Reggeons
$$
C^{R_{\pm}}(t)=C^{\pm}e^{2b^{\pm}t}, ~~C^{\pm}=C^{\pm}(0), ~~
\pm ~ \mbox{this} ~R_{+}~\mbox{or}~ R_{-}.
$$
Pomeron and Odderon
$$
C^{P}(t)=C^{P}
\left[ d_{P}e^{b_{1}^{P}}+(1-d_{P})e^{b_{2}^{P}t} \right] ,~~
C^{O}(t)=C^{O}
\left[ d_{O}e^{b_{1}^{O}}+(1-d_{O})e^{b_{2}^{O}t} \right] .
$$
$$
F^{P}(s,t)=-2m_{p}^2
\underbrace{C^{P}\left[ d_{P}e^{b_{1}^{P}}+(1-d_{P})e^{b_{2}^{P}t} \right]}_{C^{P}(t)}
(-iz_{t})^{\alpha_{P}(0)+\alpha ` _{P}t}
$$
$$
F^{O}(s,t)=-2im_{p}^2
\underbrace{
C^{O}\left[ d_{O}e^{b_{1}^{O}}+(1-d_{O})e^{b_{2}^{O}t} \right]}_{C^{O}(t)}
(-iz_{t})^{\alpha_{O}(0)+\alpha ` _{O}t}.
$$
$$
F^{R_{+}}(s,t)=-2m_{p}^2 C^{+}e^{2b^{+}t}
(-iz_{t})^{\alpha_{+}(0)+\alpha ` _{+}t}, ~~
F^{R_{-}}(s,t)=-2im_{p}^2 C^{-}e^{2b^{-}t}
(-iz_{t})^{\alpha_{-}(0)+\alpha ` _{-}t},
$$
Further on:
$$
B_{1}^{P}=b_{1}^{P}+\alpha ` _{P}\left[\ln (z_{t})-i\frac{\pi}{2}\right] ,~~
B_{2}^{P}=b_{2}^{P}+\alpha ` _{P}\left[\ln (z_{t})-i\frac{\pi}{2}\right] ,
$$
$$
B_{1}^{O}=b_{1}^{Od}+\alpha ` _{O}\left[\ln (z_{t})-i\frac{\pi}{2}\right] ,~~
B_{2}^{P}=b_{2}^{Od}+\alpha ` _{O}\left[\ln (z_{t})-i\frac{\pi}{2}\right] ,
$$
$$
F^{PP}(s,t)=\frac{-i(z C^{P})^2}{16\pi s (\hbar c)^2 \sqrt{1-4m_{p}^2 /s}}
\left[
\frac{d_{P}^2}{2B_{1}^{P}}e^{tB_{1}^{P}/2}+
\frac{d_{P}(1-d_{P})}{B_{1}^{P}+B_{2}^{P}}
e^{t\frac{B_{1}^{P}B_{2}^{P}}{B_{1}^{P}+B_{2}^{P}}}+
\frac{(1-d_{P})^2}{2B_{2}^{P}}e^{tB_{2}^{P}/2}
\right] ,
$$
$$
F^{PO}(s,t)=\frac{(z^2 C^{P}C^{O})}{8\pi s (\hbar c)^2 \sqrt{1-4m_{p}^2 /s}}
\left[
\frac{d_{P}d_{O}}{B_{1}^{P}+B_{1}^{O}}
e^{t\frac{B_{1}^{P}B_{1}^{O}}{B_{1}^{P}+B_{1}^{O}}}+
\frac{d_{P}(1-d_{O})}{B_{1}^{P}+B_{2}^{O}}
e^{t\frac{B_{1}^{P}B_{2}^{O}}{B_{1}^{P}+B_{2}^{O}}}+\right.
~~~~~~~~~~~~~~~ 
$$
$$
~~~~~~~~~~~~~~~~~~~~~~~~~~~~~~~~~~~~~~~~~~~~~~~~~~
\left. +\frac{d_{O}(1-d_{P})}{B_{2}^{P}+B_{1}^{O}}
e^{t\frac{B_{2}^{P}B_{1}^{O}}{B_{2}^{P}+B_{1}^{O}}}+
\frac{(1-d_{P})(1-d_{O})}{B_{2}^{P}+B_{2}^{O}}
e^{t\frac{B_{2}^{P}B_{2}^{O}}{B_{2}^{P}+B_{2}^{O}}}+
\right] ,
$$
$$
F^{OO}(s,t)=\frac{-i(z C^{O})^2}{16\pi s (\hbar c)^2 \sqrt{1-4m_{p}^2 /s}}
\left[
\frac{d_{O}^2}{2B_{1}^{O}}e^{tB_{1}^{O}/2}+
\frac{2d_{O}(1-d_{O})}{B_{1}^{O}+B_{2}^{O}}
e^{t\frac{B_{1}^{O}B_{2}^{O}}{B_{1}^{O}+B_{2}^{O}}}+
\frac{(1-d_{O})^2}{2B_{2}^{O}}e^{tB_{2}^{O}/2}
\right] ,
$$
$$
P^{H}(t)=i\frac{C^{PH}}{(1-t/t_{PH})^4},~~~
P^{O}(t)=i\frac{C^{OH}}{(1-t/t_{OH})^4}.
$$
Now \footnote{$m_{\pi}= 0.134977$ GeV is the mass of the neutral pion}:
$$
q_{+}=2m_{\pi}-\sqrt{4m_{\pi}^2-t}, ~~~
q_{-}=3m_{\pi}-\sqrt{9m_{\pi}^2-t},
$$
$$
\Phi_{H,1}(t)=b_{1}^H q_{+},~~~
\Phi_{H,2}(t)=b_{2}^H q_{+},~~~
\Phi_{H,3}(t)=b_{3}^H q_{+},
$$
$$
\Phi_{O,1}(t)=b_{1}^O q_{-},~~~
\Phi_{O,2}(t)=b_{2}^O q_{-},~~~
\Phi_{O,3}(t)=b_{3}^O q_{-} ,
$$
Notation: $\tau=\sqrt{-t/t_0},~~t_0 = 1~\mbox{GeV}^2$.
$$
F^{H}(s,t)=i\overbrace{2m_{p}^2 z_{t}}^{z(s,t)}
\left[H_{1}\zeta^2\frac{2J_{1}(r_{+}\tau \zeta )}{r_{+}\tau \zeta}
\Phi^2_{H,1}(t)+
H_{2}\zeta^2\frac{\sin (r_{+}\tau \zeta )}{r_{+}\tau \zeta}
\Phi^2_{H,2}(t)+
H_{3} J_{0}(r_{+}\tau \zeta) \Phi^2_{H,3}(t)\right]
$$
$$
F^{MO}(s,t)=i\underbrace{2m_{p}^2 z_{t}}_{z(s,t)}
\left[O_{1}\zeta^2\frac{2J_{1}(r_{-}\tau \zeta )}{r_{-}\tau \zeta}
\Phi^2_{O,1}(t)+
O_{2}\zeta^2\frac{\sin (r_{-}\tau \zeta )}{r_{-}\tau \zeta}
\Phi^2_{O,2}(t)+
O_{3} J_{0}(r_{-}\tau \zeta) \Phi^2_{O,3}(t)\right]
$$
$$
\mathbf{F^{+}(s,t)=F^{H}(s,t)+F^{P}(s,t)+F^{R_{+}}(s,t)+F^{PP}(s,t)+
F^{OO}(s,t)+P^{H}(t)},
$$
$$
\mathbf{F^{-}(s,t)=F^{MO}(s,t)+F^{O}(s,t)+F^{R_{-}}(s,t)+F^{PO}(s,t)+
P^{O}(t)}.
~~~~~~~~~~~~~~ 
$$

Finally we have for nuclear amplitudes:\vspace{-2.1mm}
$$
F_{pp}^{N}=F^{+}+F^{-},~~~F_{{\bar p}p}^{N}=F^{+}-F^{-} .\vspace{-2.1mm}
$$
\subsubsection*{Coulomb amplitude $F^C$ \footnote{Note that in a number of works (see, for example, \cite{Pet}) it has been proven that the use of such a parametrization, strictly speaking, is incorrect.
However, within the limits of the accuracy adopted here, this does not seem to be significant.}\vspace{-2.1mm}}

$$
B_{pp}(s)=\frac{\sigma^{tot}_{pp}}{4\pi (\hbar c)^2},~~~
B_{{\bar p}p}(s)=\frac{\sigma^{tot}_{{\bar p}p}}{4\pi (\hbar c)^2}
.\vspace{-2.1mm}
$$
$$
\Phi_{pp}(s,t)=-\ln\left[-\frac{t}{2}
\left(B_{pp}(s)+\frac{8\pi}{\Lambda^2}\right)\right]
-\gamma-\frac{4t}{\Lambda^2}\ln\left(-\frac{4t}{\Lambda^2}\right)-
\frac{2t}{\Lambda^2} ,\vspace{-2.1mm}
$$
$$
\Phi_{{\bar p}p}(s,t)=-\ln\left[-\frac{t}{2}
\left(B_{{\bar p}p}(s)+\frac{8\pi}{\Lambda^2}\right)\right]
-\gamma-\frac{4t}{\Lambda^2}\ln\left(-\frac{4t}{\Lambda^2}\right)-
\frac{2t}{\Lambda^2} ,\vspace{-2.1mm}
$$
where $\Lambda$ = $\sqrt{0.71}$ GeV,
$\alpha=7.2973525693\cdot 10^{-3}\cong 1/137$ is the fine structure constant\vspace{-1.1mm}
$$
\mathbf{
F^{C}_{pp}=e^{i\alpha\Phi_{pp}(s,t)}
\frac{8\pi (\hbar c)^2 s\alpha}
{t\left(\frac{4m_{p}^2-2.79 t}{4m_{p}^2-t}
\right)^{2}\left( 1-\frac{t}{\Lambda^2}
\right)^{4}} ,~~~
F^{C}_{{\bar p}p}=-e^{-i\alpha\Phi_{pp}(s,t)}
\frac{8\pi (\hbar c)^2 s\alpha}
{t\left(\frac{4m_{p}^2-2.79 t}{4m_{p}^2-t}
\right)^{2}\left( 1-\frac{t}{\Lambda^2}
\right)^{4}}} .
$$
Fitting parameters:\vspace{1.1mm}

$H_1, H_2, H_3, b_1^{H}, b_2^{H}, b_3^{H}, r_{+}, b_1^{P}, b_2^{P}, \alpha'_P, C^P, B^{+},
C^{+}, \alpha'_{+}, \alpha_{+}(0), C^{PH}, t^{PH}, d_P, \alpha_P(0)$,\vspace{1.1mm}

$O_1, O_2, O_3, b_1^{O}, b_2^{O}, b_3^{O}, r_{-}, b_1^{Od}, b_2^{Od}, \alpha'_O, C^O, B^{-},
C^{-}, \alpha'_{-}, \alpha_{-}(0), C^{OH}, t^{OH}, d_O, \alpha_O(0)$.


\end{document}